\documentclass[prb,superscriptaddress,preprint,showpacs,showkeys]{revtex4}

\usepackage{graphicx}
\usepackage{bm}
\usepackage{amsmath}
\usepackage{amssymb}

\begin{document}

\title{Electrostatic deflections of cantilevered metallic carbon nanotubes via charge-dipole model}

\author{Zhao Wang}
\email{wzzhao@yahoo.fr}
\affiliation{Institute UTINAM, UMR 6213, University of Franche-Comt\'{e}, 25030 Besan\c {c}on Cedex, France.}
\author{Michel Devel}
\affiliation{Institute UTINAM, UMR 6213, University of Franche-Comt\'{e}, 25030 Besan\c {c}on Cedex, France.}

\begin{abstract}
We compute electrostatic fields induced deformations of cantilevered finite-length metallic carbon nanotubes, using an energy minimization method based on a charge-dipole moment interaction potential combined with an empirical many-body potential. The influence of field strength, field direction and tube geometry on the electrostatic deflection is investigated for both single and double walled tubes. These results could apply to nanoelectromechanical devices based on cantilevered carbon nanotubes.
\end{abstract}


\maketitle

\section{Introduction}
\label{intro}

Cantilevered CNTs can be used as key elements in nanoelectromechanical systems (NEMS) such as nanorelays,\cite{kinaret-03,lee-04} nanoswitches,\cite{jang-05} nanotweezers\cite{akita-01} and feedback device\cite{ke-04} which are designed for memory, sensing or actuation uses. The electric field induced movements and deformations are key characteristics for these applications, as well as for CNTs' fabrication\cite{srivastava-01,joselevich-02,li-04} and separation.\cite{krupke-03} Indeed, electric field induced deflection\cite{Poncharal-99,wei-01}, alignment\cite{chen-01,kumar-04} and microstructure change\cite{bao-01} of CNTs have been observed in experiments. Compared to the semiconducting CNTs, the metallic ones can generally be more sensitive to the presence of external electric fields due to their free charge distribution and higher polarizability.\cite{joselevich-02} They can therefore be expected to play a more important role in NEMS. 

In this work, motionless equilibrium forms of CNTs in electric fields are computed by the method of energy minimization using the algorithm of conjugated gradient. The total potential energy of the system is the sum of an induced electrostatic potential and a many-body the interatomic potential. In one of our previous studies, this method was used with a regularized dipole-only model\cite{langlet-06} to calculate the induced electrostatic potential of semiconducting single-walled CNTs (SWCNTs).\cite{zhaowang-06-01} However, the change of the charge distribution of metallic CNTs cannot be properly described by this model. Thus, a regularized charge-dipole model parameterized for fullerenes and metallic CNTs,\cite{mayer-05-01,mayer-06-01,mayer-07-01} is used in this work. 

The interatomic potential is computed using the adaptive interatomic reactive empirical bond order (AIREBO) potential.\cite{stuart-00} This potential is an evolution of a many-body chemical pseudopotential model (REBO) parameterized by Brenner\cite{Brenner-90} for conjugated hydrocarbons, which has been widely used in theoretical studies on mechanical and thermal properties of CNTs. Polarization effects may change the strength of bonds, as discussed in Ref.\cite{guo2003}. Indeed, Figs 6 and 7 of this reference show that the relation between the deformations is significantly changed in fields with respective strengths E = 32 V/nm and E = 40 V/nm. However, since the fields used in our paper (0.1 to 3.0 V/nm) are at least 10 times weaker than their fields, we think it is a reasonable approximation to neglect the influence of the electric field on the parameters of the AIREBO potential and use a separate potential energy to take into account the interaction with the field.
 
The details about the models will be presented in sec.\,II. The results for both SWCNTs and double-walled CNTs (DWCNTs) are shown and discussed in sec.\,III. We draw conclusions in sec.\,IV.


\section{Physico-chemical model}
At the beginning of calculation, open-ended tubes with zero net charge and zero permanent dipole moment are fixed at one of their two ends on an substrate which is supposed to be insulating in order to allow us to neglect transfer of charges from the nanotube to the substrate. Each atom is associated with both an induced dipole $\bm{p_i}$ and a quantity of induced charge $q_i$ when the tube is submitted to an electric field. The total energy of this system $U^{tot}$ is the sum of the induced electrostatic energy $U^{elec}$ and the interatomic potential $U^{p}$ : $U^{tot}=U^{elec}+U^{p}$, in which $U^{elec}$ can be written as follows:

\begin{multline}
\label{eq:2}
U^{elec}=\sum_{i=1}^N{q_i(\chi_i+V_i)}-\sum_{i=1}^N{\bm{p}_i\cdot\bm{E_i}}+
\frac{1}{2}\sum_{i=1}^N{\sum_{\substack{j=1}}^N{q_i T^{i,j}_{q-q} q_j}}\\-\sum_{i=1}^N{\sum_{\substack{j=1}}^N{\bm{p}_i \cdot \bm{T}^{i,j}_{p-q} q_j}}
-\frac{1}{2}\sum_{i=1}^N{\sum_{\substack{j=1}}^N{\bm{p}_i \cdot \bm{T}^{i,j}_{p-p} \cdot \bm{p}_j}}
\end{multline}

where $N$ is the total number of atoms, $\chi_i$ stands for the electron affinity of atom $i$, $V_i$ is the external potential, $\bm{E_i}$ represents the external electric field. $T$ and $\bm{T}$ stand for the vacuum electrostatic propagators regularized by a Gaussian distribution in order to avoid the divergence problem when two atoms are too close to each other. They can be written as $T^{i,j}_{q-q}=\frac{1}{4\pi\epsilon_0}\frac{\texttt{erf}\left[r_{i,j}/\left(\sqrt{2}R\right)\right]}{r_{i,j}}$,  $\bm{T}^{i,j}_{p-q}=-\nabla_{\bm{r}_i} T^{i,j}_{q-q}$ and $\bm{T}^{i,j}_{p-p}=-\nabla_{\bm{r}_j} \otimes \nabla_{\bm{r}_i} T^{i,j}_{q-q}$, where $\bm{r}_i$ represents the coordinate of atom $i$, $r_{i,j}$ stands for the distance between atom $i$ and atom $j$, and $R$ is the width of the Gaussian distribution of charge. The value of $R$ used in this work is about $0.06862$\,nm, which was fitted to reproduce the polarizability of metallic tubes\,\cite{mayer-06-01}.

Taking $\lim{r_{i,j}\rightarrow 0}$, we obtain the self energy terms (when $i=j$ in Eq. 2) as follows: 

\begin{subequations} 
\label{eq:3}
\begin{align} 
&\frac{1}{2}q_{i}T^{i,i}_{q-q}q_{i} = \frac{1}{4\pi\epsilon_0}\frac{\sqrt{2/\pi}}{R}\frac{q^2_i}{2}\\
&\bm{p}_{i}\cdot\bm{T}^{i,i}_{p-q}q_{i} = 0\\
&\frac{1}{2}\bm{p}_{i}\cdot\bm{T}^{i,i}_{p-p}\cdot\bm{p}_{i} =\frac{1}{2}\bm{p}_{i}\cdot\alpha^{-1}_{i}\cdot\bm{p}_{i}
\end{align}
\end{subequations}

where $\alpha_{i}$ stands for the polarizability of atom $i$. 

\begin{figure}[ht]
\centerline{\includegraphics[width=14cm]{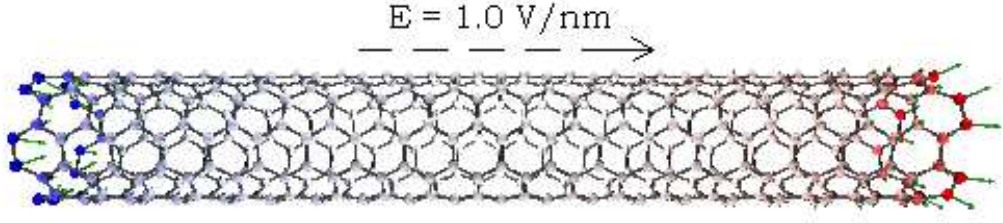}}
\caption{\label{fig:0}
(Color online) Induced dipoles and charges on a open-ended metallic (5,5) SWCNTs (L=4.8nm) subjected to a horizontal electric field $E=1.0$\,V/nm. The positive charges move to the right side and the negative ones move to the left (the color scaling is proportional to the density of charge). The green vectors stand for the dipoles. The maximal amplitudes of these charges and dipoles are about $0.34$\,unit of \textit{e} and $3.2 e \cdot nm$, respectively.}
\end{figure}

The distributions of dipoles $\left\{\bm{p}_i\right\}$ and charges $\left\{q_i\right\}$ is determined by the fact that the static equilibrium state of these distributions should correspond to the minimum value of $U^{elec}$. Thus, by requiring that the partial derivatives of the total electrostatic energy with respect to the $3 \times N$ components of the dipoles and $N$ components of the charges should be zero within Eq.\,\ref{eq:2}, and taking account of Eqs.\,\ref{eq:3}, we obtain $\left\{\bm{p}_i\right\}$ and $\left\{q_i\right\}$ by solving $N$ linear vectorial equations and $N$ linear scalar equations as follows: 
\begin{equation}
\label{eq:4}
\forall{i=1,...,N}\quad\left\{ \begin{array}{ll}
\sum\limits_{j=1}^N{\bm{T}^{i,j}_{p-p}\otimes\bm{p}_{j}}+\sum\limits_{j=1}^N{\bm{T}^{i,j}_{p-q}q_{j}}=-\bm{E_i}& \\
\sum\limits_{j=1}^N{\bm{T}^{i,j}_{p-q}\cdot\bm{p}_{j}}+\sum\limits_{j=1}^N{T^{i,j}_{q-q}q_{j}}=-(\chi_i+V_i)& \end{array}\right.
\end{equation}

In Fig.\,\ref{fig:0}, we show the distribution of the dipoles and charges induced by an electric field on a metallic SWCNT.
 
The interatomic potential $U^p$ is computed using the AIREBO potential function \cite{stuart-00}. This potential is an extension of Brenner's second generation potential (Brenner \textit{et al}\cite{brenner-02}) and includes long-range atomic interactions and single bond torsional interactions. In this type of potential, the total interatomic potential energy is the sum of individual pair interactions containing a many-body bond order function as follows:

\begin{equation}
\label{eq:5}
U^p=\frac{1}{2}\sum\limits_{i=1}^N{\sum\limits_{\substack{j=1 \\ j\ne i} }^N{\left[ 
\begin{array}{l}
\varphi^R\left(r_{i,j}\right)
-b_{i,j}\varphi^A\left(r_{i,j}\right)
+\varphi^{LJ}\left(r_{i,j}\right)
+\sum\limits_{\substack{k=1 \\ k\ne i,j} }^N
{\sum\limits_{\substack{\ell=1 \\ \ell\ne i,j,k}}^N
{\varphi_{kij\ell}^{tor}}}
\end {array}
\right]} }
\end{equation}

where $\varphi^R$ and $\varphi^A$ are the interatomic repulsion and attraction terms between valence electrons, respectively, for bound atoms. The bond order function $b_{i,j}$ provides the many body effects by depending on the local atomic environment of atoms $i$ and $j$. The long-range interactions are included by adding $\varphi^{LJ}$, a parameterized \textit{Lennard-Jones 12-6} potential term. $\varphi^{tor}$ represents the torsional interactions.  

Energy optimization is performed to obtain the motionless equilibrium configurations of the atoms using the method of conjugated gradient\,\cite{nrf77}. We note that during this process the induced net charges and dipoles on each atom are updated at every step of the minimization procedure (which is quite time-consuming). 

\section{Results and discussion}
In this work, Cartesian coordinates are used with the $z$ axis along the principal axis of the tube. The open-ended tubes are fixed at one end to an insulating substrate and relaxed in free space before being submitted to a homogeneous electric field. All applied fields are parallel to the $y$-$z$ plane. The field strengths are between 0.1 and 3.0 V/nm for the metallic tubes and between 0.1 and 1.0 V/nm for semiconducting tubes. We note that Li~\textit{et al.} showed that a semiconductor-metal transition takes place in a (16,0) CNT, when electric fields reach about 3.0-4.0 V/nm\,\cite{Li2006}. Furthermore, in actual experiments, the field strengths needed to get comparable deflections are much weaker than those used here, since we use tubes at least 100 times shorter than in experiment and, as shown hereafter, the longer the tube, the weaker the field needed to get a given deflection. This is the same as in field emission experiments in which the shorter the tube, the stronger the field strength needed to produce a given field emission intensity, owing to the decrease of the tip effect on the field enhancement factor (see e.g. Fig.~3 of Ref.~\cite{Jo2003}).

\begin{figure}[ht]
\centerline{\includegraphics[width=14cm]{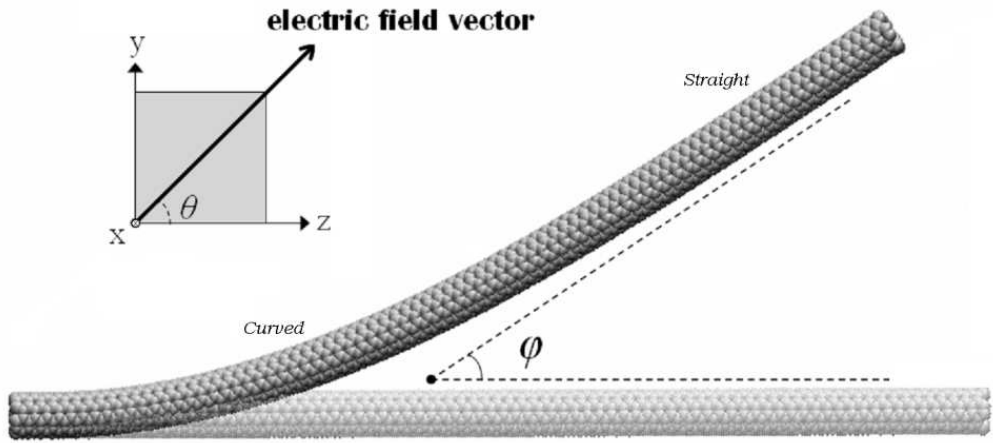}}
\caption{\label{fig:1}
Electrostatic deflection of a SWCNTs (5,5), tube length $L\approx19.8$\,nm, field angle $\theta=45^\circ$, field strength $E=\left|\bm{E}\right|=0.775$\,V/nm.}
\end{figure} 

Fig.\,\ref{fig:1} shows the equilibrium position of a SWCNT in a uniform electric field. It can be seen from this figure that the tube is only curved at the part close to the fixed end. We find that its right side part remains straight and that it is slightly compressed by electrostrictive effects\cite{guo-03-02} comparing its average bond lengths before and after deflection. We note that in real experiments, the tube would have thermal vibrations around this equilibrium position,\cite{treacy-96,zhang-01} and that this deformation is generally reversible.\cite{wei-01} Furthermore, the bending of the fixed end would generally not lead to important changes of the tube conductivity\,\cite{Rochefort1999,Nardelli1999}.

As shown in Fig.\,\ref{fig:1}, the field angle $\theta$ is defined as the angle between the field direction and the $z$ axis, the deformation angle $\varphi$ is defined as the angle between the neutral axis of the deformed CNTs at the free end and the $z$ axis. 

\begin{figure}[ht]
\centerline{\includegraphics[width=11cm]{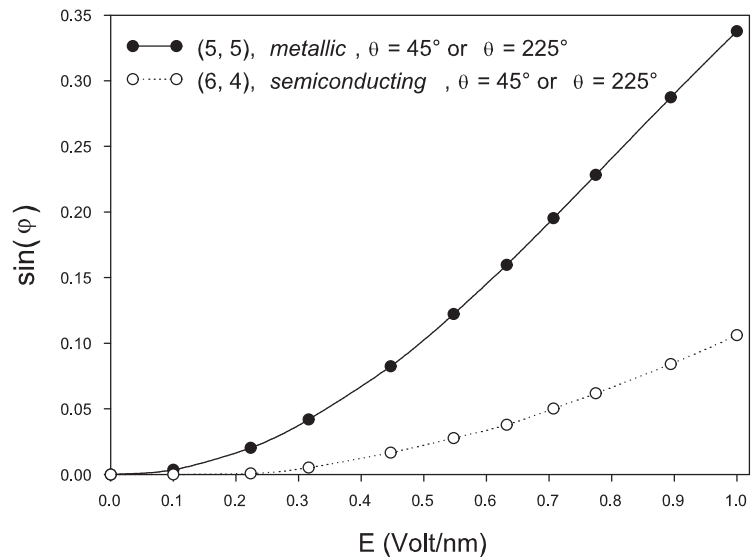}}
\caption{\label{fig:2}
$\sin(\varphi)$ \textit{versus} $E$ for two SWCNTs: a metallic one (5,5),  $L\approx13.16$\,nm, radius $R=0.34$\,nm), and a semiconducting one (6,4) with the same $L$ and $R$. The fields are applied in both $\theta=\pi/4$ and $\theta=5\pi/4$ (in opposite direction).}
\end{figure}

Fig.\,\ref{fig:2} shows the relation between the external fields and the deformation angles $\varphi$ for two SWCNTs. The deflection of the semiconducting tube is calculated using the dipole-only model with parameters given in Ref.\cite{langlet-06}, while we use the charge-dipole model with parameters from Ref.\cite{mayer-06-01} for the metallic one. As expected, it can be seen that the deflection of the (5,5) tube is much larger than that of the (6,4) semiconducting one for a given electric field, and that the tube deflection is the same no matter whether the field direction is reversed. Furthermore, we note that the form of the curves of $\sin(\varphi)$ \textit{versus} $E$ is in a  qualitative agreement with the results of the experiment of Poncharal \textit{et al} (Fig.\,1 in Ref.\cite{Poncharal-99}), and we find $\sin(\varphi) \propto E^{2}$ when the deflection is relatively small ($\sin(\varphi)<0.15$) for both of these two CNTs. 

\begin{figure}[ht]
\centerline{\includegraphics[width=11cm]{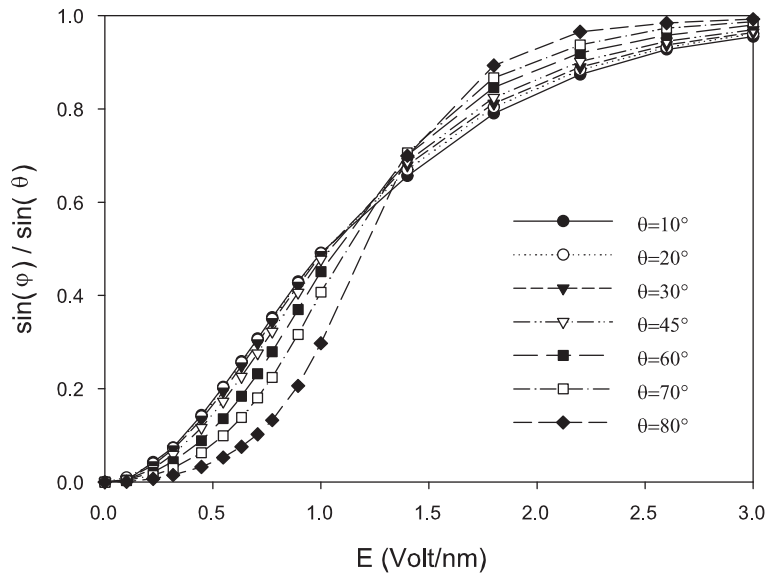}}
\caption{\label{fig:3}
$\sin(\varphi)/\sin(\theta)$ \textit{versus} $E$ for a metallic tube (5,5), $L\approx13.16$\,nm.}
\end{figure}

For higher field strength, the alignment ratio is defined as $\sin(\varphi)/\sin(\theta)$. It is calculated for several field directions and plotted in Fig.\,\ref{fig:3}. It stands for the relative deformation to the field direction and attains its maximum value $1$ once the tube is well aligned to the field. We can see that, when the value of $E$ remains small ($<1.4$\,V/nm), the alignment ratio is larger for the smaller field angles $\theta$. On the other hand, this tube can be more efficiently aligned to the field direction in stronger fields for larger field angles. No deflection is found when the field is perfectly perpendicular to the tube axis, because the induced molecular dipole is already aligned to the field. However, we note that this case can hardly happen in realistic experimental condition due to the thermal vibration of the tube and the fact that generally the CNTs are more or less naturally curved due to the presence of defects.  

\begin{figure}[ht]
\centerline{\includegraphics[width=10cm]{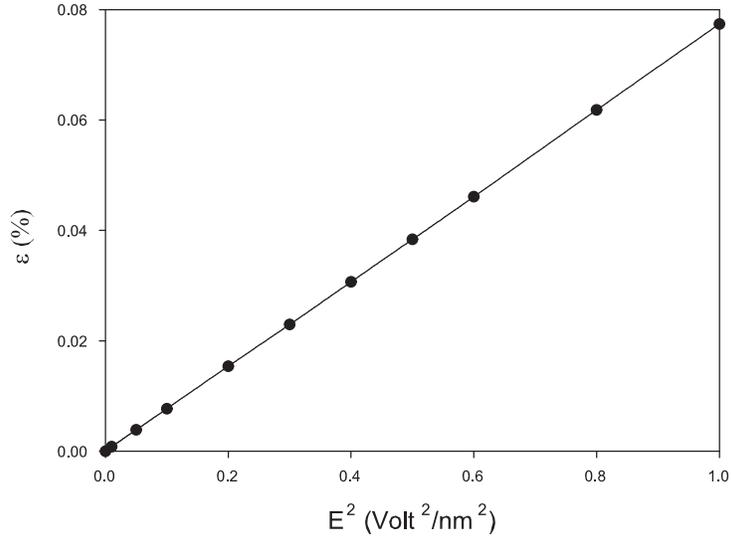}}
\caption{\label{fig:4}
Axial strain $\varepsilon=\Delta{L}/L$ (\%) \textit{versus} $E^2$ for a metallic tube (5,5)($L\approx13.16$\,nm), when the electric fields are applied parallel to the tube axis.}
\end{figure}

As expected, there is no electrostatic deflection found when $\theta=0$. Since the induced molecular dipole of the tube is already aligned to the direction of the field, the total induced torque acting on the tube is therefore zero. Nevertheless, slight electrostriction effects are found in the axial direction of the tube. The electrostrictive deformation $\varepsilon=\Delta{L}/L$ is plotted in Fig.\,\ref{fig:4} \textit{versus} the square of field strength. It can be seen that $\varepsilon$ is proportional to $E^2$ for these field strengths. 
This numerical experiment also allowed us to estimate the nanotube Young's modulus ($Y$) directly by the stress over strain ratio, since the axial external electrostatic force acting on the tube can be directly computed in our program. Using the commonly adopted wall thickness value of $0.34$~nm, we find that $Y$ is about 0.95 TPa, which is in good agreement with the average of the values found in the literature for that thickness (see e.g. section 2.1 of the recent review by Coleman et al.\cite{coleman-carbon-2006}).

\begin{figure}[ht]
\centerline{\includegraphics[width=14cm]{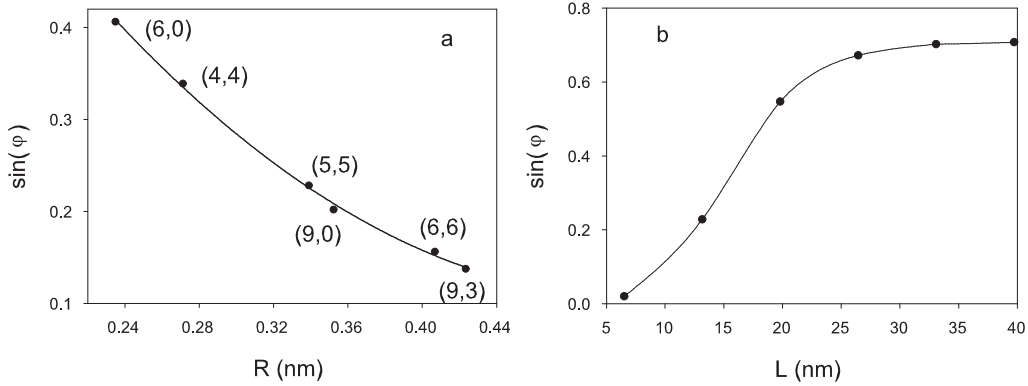}}
\caption{\label{fig:5}
In an external electric field $E=0.775$\,V/nm, $\theta=45^\circ$, a: $\sin(\varphi)$ \textit{versus} the radii $R$ of six metallic tubes with the same length $L\approx 13.2$\,nm; b: $\sin(\varphi)$ \textit{versus} the length of six (5,5) CNTs $L \approx 6.52$, $13.16$, $19.8$, $26.44$, $33.08$ and $39.72$\,nm.}
\end{figure}

Turning back to the question of electrostatic deflection, we also study tube geometry effects. Fig.\,\ref{fig:5}\textit{a} shows  $\sin(\varphi)$ \textit{versus} the radius $R$ for several metallic CNTs with the same length. It can be seen that the bigger the tube radius, the smaller the induced deflection. It is well known that the polarization effects are more important when the tube is bigger. However, at the same time, the tube becomes harder to be bent due to the increase of the moment of inertia of its cross section. From our results, it is obvious that the later effect plays a more important role. Note also that the deflection of zigzag tubes is slightly smaller than that of chiral and armchair ones, due to their larger elastic moduli\cite{zhaowang-06}. This curve of $\sin(\varphi)$ can be fitted as: $\sin(\varphi)=1/(2.5886R^2-0.3839)$. Hence $\sin(\varphi)$ is roughly proportional to $1/R^{2}$ for this electric field.

Fig.\,\ref{fig:5}\textit{b} shows the relation between the deflection and the tube length. We can see that the deflection increases significantly with the increase of tube length when $\varphi$ remains much smaller than $\theta$. Then, it reaches a plateau. It can be seen that the form of the curve of $\sin(\varphi)$ \textit{versus} $L$ is very similar to that of $\sin(\varphi)/\sin(\theta)$ \textit{versus} $E$. This is probably because $L$ and $E$ play two similar roles in the total induced torque $T=\beta(L) E^{2} \sin(\varphi-\theta) \cos(\varphi-\theta)$(where $\beta$ is the molecular polarizability of CNTs).\cite{kozinsky-06} Hence, considering that the lengths of the CNTs studied in previous experimental works are in the range from hundreds of nanometers to some micrometers, the required field strength can be much lower than the fields used in this paper for a given deflection angle. Furthermore, for letting the readers conveniently find the values of $\sin(\varphi)$ in Fig.\,\ref{fig:5}\textit{b}, we give the best fitting function of this curve as  $\sin(\varphi)= \sin(\theta)/(1+(L/15.3010)^{-5})$.

\begin{figure}[ht]
\centerline{\includegraphics[width=14cm]{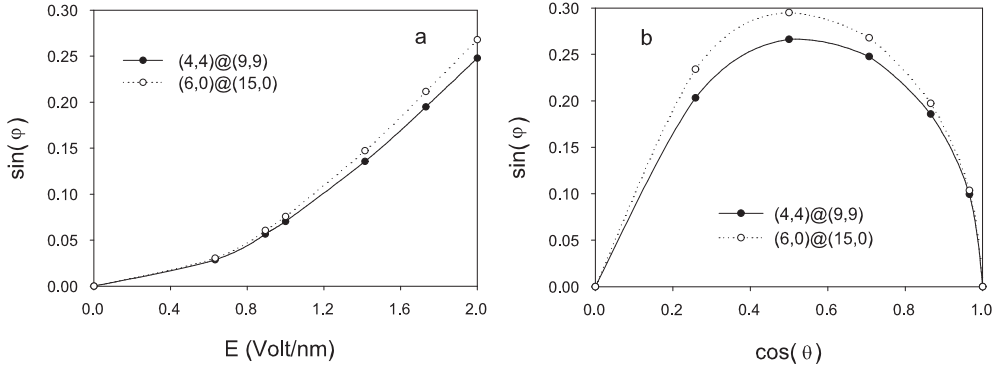}}
\caption{\label{fig:6}
For two metallic DWCNTs: (4,4)@(9,9) with $L\approx12.2$\,nm, $R^{inner}\approx0.27$\,nm, $R^{outer}\approx0.61$\,nm and (6,0)@(15,0) with $L\approx12.2$\,nm, $R^{inner}\approx0.23$\,nm, $R^{outer}\approx0.59$\,nm. a: $\sin(\varphi)$ \textit{versus} $E$, $\theta=45^\circ$. b: $\sin(\varphi)$ \textit{versus} $cos(\theta)$, $E=2$Volt/nm.}
\end{figure}

Fig.\,\ref{fig:6}\textit{a} shows the relation between the deflection and the field strength for two DWCNTs. It is found that the deflection of DWCNTs remains small even in strong electric fields. For a metallic cylinder, the screening factor is very high thus the inner layer is almost completely screened. On the other hand, their important effective cross sections increase with increasing layer number. Thus, a MWCNT can be much harder to be bent by the electric field than a SWCNT with the same radius. $\sin(\varphi)$ is also plotted in Fig.\,\ref{fig:6}\textit{b} for several field directions. We can see the DWCNTs can be most efficiently bent at $\theta=60^\circ$, like SWCNTs, for this field intensity. This value can be biased towards 90 degrees because the axial polarizability of CNTs is always greater than the radial one.

\section{Conclusion}
In this paper, we investigate the mechanisms of the electrostatic deflection of cantilevered metallic SWCNTs and DWCNTs. The equilibrium positions of CNTs in electric fields are calculated. The metallic CNTs are much easier to be deflected than semiconducting ones. The deflection is not changed by reversing the field direction. The curve of alignment ratio \textit{versus} field strength is found to change with field directions. The deflection is found to decrease with the increase of the tube radius; conversely, it increases when the tube is longer. The multi-walled metallic CNTs are found to be much harder to be bent in electric fields than SWCNTs. Furthermore, we find that the electrostrictive deformation of SWCNTs is proportional to the square of field strength. Uniform external fields are applied as a theoretical simplification. However, our scheme is able to deal with inhomogeneous fields such as those from real experiments.

We believe that this paper could help developing a better understanding of recently designed NEMS based on cantilevered CNTs.  We also wish that these results can be useful to open the path to some new nanoelectromechanical devices. 

\begin{acknowledgments}
This work was done as parts of the CNRS GDR-E Nb 2756. Z. W. acknowledges the support received from the region of Franche-Comt\'{e} (grant 060914-10). 
\end{acknowledgments}


\begin{thebibliography}{34}
\expandafter\ifx\csname natexlab\endcsname\relax\def\natexlab#1{#1}\fi
\expandafter\ifx\csname bibnamefont\endcsname\relax
  \def\bibnamefont#1{#1}\fi
\expandafter\ifx\csname bibfnamefont\endcsname\relax
  \def\bibfnamefont#1{#1}\fi
\expandafter\ifx\csname citenamefont\endcsname\relax
  \def\citenamefont#1{#1}\fi
\expandafter\ifx\csname url\endcsname\relax
  \def\url#1{\texttt{#1}}\fi
\expandafter\ifx\csname urlprefix\endcsname\relax\def\urlprefix{URL }\fi
\providecommand{\bibinfo}[2]{#2}
\providecommand{\eprint}[2][]{\url{#2}}

\bibitem[{\citenamefont{Kinaret et~al.}(2003)\citenamefont{Kinaret, Nord, and
  Viefers}}]{kinaret-03}
\bibinfo{author}{\bibfnamefont{J.~M.} \bibnamefont{Kinaret}},
  \bibinfo{author}{\bibfnamefont{T.}~\bibnamefont{Nord}}, \bibnamefont{and}
  \bibinfo{author}{\bibfnamefont{S.}~\bibnamefont{Viefers}},
  \bibinfo{journal}{Appl. Phys. Lett.} \textbf{\bibinfo{volume}{82}},
  \bibinfo{pages}{1287} (\bibinfo{year}{2003}).

\bibitem[{\citenamefont{Lee et~al.}(2004)\citenamefont{Lee, Lee, Morjan, Jhang,
  Sveningsson, Nerushev, Park, and Campbell}}]{lee-04}
\bibinfo{author}{\bibfnamefont{S.~W.} \bibnamefont{Lee}},
  \bibinfo{author}{\bibfnamefont{D.~S.} \bibnamefont{Lee}},
  \bibinfo{author}{\bibfnamefont{R.~E.} \bibnamefont{Morjan}},
  \bibinfo{author}{\bibfnamefont{S.~H.} \bibnamefont{Jhang}},
  \bibinfo{author}{\bibfnamefont{M.}~\bibnamefont{Sveningsson}},
  \bibinfo{author}{\bibfnamefont{O.~A.} \bibnamefont{Nerushev}},
  \bibinfo{author}{\bibfnamefont{Y.~W.} \bibnamefont{Park}}, \bibnamefont{and}
  \bibinfo{author}{\bibfnamefont{E.~E.~B.} \bibnamefont{Campbell}},
  \bibinfo{journal}{Nano Lett.} \textbf{\bibinfo{volume}{4}},
  \bibinfo{pages}{2027} (\bibinfo{year}{2004}).

\bibitem[{\citenamefont{Jang et~al.}(2005)\citenamefont{Jang, Cha, Choi,
  Amaratunga, Kang, Hasko, Jung, and Kim}}]{jang-05}
\bibinfo{author}{\bibfnamefont{J.}~\bibnamefont{Jang}},
  \bibinfo{author}{\bibfnamefont{S.}~\bibnamefont{Cha}},
  \bibinfo{author}{\bibfnamefont{Y.}~\bibnamefont{Choi}},
  \bibinfo{author}{\bibfnamefont{G.}~\bibnamefont{Amaratunga}},
  \bibinfo{author}{\bibfnamefont{D.}~\bibnamefont{Kang}},
  \bibinfo{author}{\bibfnamefont{D.}~\bibnamefont{Hasko}},
  \bibinfo{author}{\bibfnamefont{J.}~\bibnamefont{Jung}}, \bibnamefont{and}
  \bibinfo{author}{\bibfnamefont{J.}~\bibnamefont{Kim}},
  \bibinfo{journal}{Appl. Phys. Lett.} \textbf{\bibinfo{volume}{87}},
  \bibinfo{pages}{163114} (\bibinfo{year}{2005}).

\bibitem[{\citenamefont{Akita et~al.}(2001)\citenamefont{Akita, Nakayama,
  Mizooka, Takano, Okawa, Miyatake, Yamanaka, Tsuji, and Nosaka}}]{akita-01}
\bibinfo{author}{\bibfnamefont{S.}~\bibnamefont{Akita}},
  \bibinfo{author}{\bibfnamefont{Y.}~\bibnamefont{Nakayama}},
  \bibinfo{author}{\bibfnamefont{S.}~\bibnamefont{Mizooka}},
  \bibinfo{author}{\bibfnamefont{Y.}~\bibnamefont{Takano}},
  \bibinfo{author}{\bibfnamefont{T.}~\bibnamefont{Okawa}},
  \bibinfo{author}{\bibfnamefont{Y.}~\bibnamefont{Miyatake}},
  \bibinfo{author}{\bibfnamefont{S.}~\bibnamefont{Yamanaka}},
  \bibinfo{author}{\bibfnamefont{M.}~\bibnamefont{Tsuji}}, \bibnamefont{and}
  \bibinfo{author}{\bibfnamefont{T.}~\bibnamefont{Nosaka}},
  \bibinfo{journal}{Appl. Phys. Lett.} \textbf{\bibinfo{volume}{79}},
  \bibinfo{pages}{1691} (\bibinfo{year}{2001}).

\bibitem[{\citenamefont{Ke and Espinosa}(2004)}]{ke-04}
\bibinfo{author}{\bibfnamefont{C.}~\bibnamefont{Ke}} \bibnamefont{and}
  \bibinfo{author}{\bibfnamefont{H.}~\bibnamefont{Espinosa}},
  \bibinfo{journal}{Appl. Phys. Lett.} \textbf{\bibinfo{volume}{85}},
  \bibinfo{pages}{681} (\bibinfo{year}{2004}).

\bibitem[{\citenamefont{Srivastava et~al.}(2001)\citenamefont{Srivastava,
  Srivastava, and Srivastava}}]{srivastava-01}
\bibinfo{author}{\bibfnamefont{A.}~\bibnamefont{Srivastava}},
  \bibinfo{author}{\bibfnamefont{A.~K.} \bibnamefont{Srivastava}},
  \bibnamefont{and} \bibinfo{author}{\bibfnamefont{O.~N.}
  \bibnamefont{Srivastava}}, \bibinfo{journal}{Carbon}
  \textbf{\bibinfo{volume}{39}}, \bibinfo{pages}{201} (\bibinfo{year}{2001}).

\bibitem[{\citenamefont{Joselevich and Lieber}(2002)}]{joselevich-02}
\bibinfo{author}{\bibfnamefont{E.}~\bibnamefont{Joselevich}} \bibnamefont{and}
  \bibinfo{author}{\bibfnamefont{C.~M.} \bibnamefont{Lieber}},
  \bibinfo{journal}{Nano Lett.} \textbf{\bibinfo{volume}{2}},
  \bibinfo{pages}{1137} (\bibinfo{year}{2002}).

\bibitem[{\citenamefont{Li et~al.}(2004)\citenamefont{Li, Zhang, Yang, and
  Tian}}]{li-04}
\bibinfo{author}{\bibfnamefont{J.}~\bibnamefont{Li}},
  \bibinfo{author}{\bibfnamefont{Q.}~\bibnamefont{Zhang}},
  \bibinfo{author}{\bibfnamefont{D.}~\bibnamefont{Yang}}, \bibnamefont{and}
  \bibinfo{author}{\bibfnamefont{J.}~\bibnamefont{Tian}},
  \bibinfo{journal}{Carbon} \textbf{\bibinfo{volume}{42}},
  \bibinfo{pages}{2263} (\bibinfo{year}{2004}).

\bibitem[{\citenamefont{Krupke et~al.}(2003)\citenamefont{Krupke, Hennrich,
  Lohneysen, and Kappes}}]{krupke-03}
\bibinfo{author}{\bibfnamefont{R.}~\bibnamefont{Krupke}},
  \bibinfo{author}{\bibfnamefont{F.}~\bibnamefont{Hennrich}},
  \bibinfo{author}{\bibfnamefont{H.~V.} \bibnamefont{Lohneysen}},
  \bibnamefont{and} \bibinfo{author}{\bibfnamefont{M.~M.}
  \bibnamefont{Kappes}}, \bibinfo{journal}{Science}
  \textbf{\bibinfo{volume}{301}}, \bibinfo{pages}{344} (\bibinfo{year}{2003}).

\bibitem[{\citenamefont{Poncharal et~al.}(1999)\citenamefont{Poncharal, Wang,
  Ugarte, and De~Heer}}]{Poncharal-99}
\bibinfo{author}{\bibfnamefont{P.}~\bibnamefont{Poncharal}},
  \bibinfo{author}{\bibfnamefont{Z.~L.} \bibnamefont{Wang}},
  \bibinfo{author}{\bibfnamefont{D.}~\bibnamefont{Ugarte}}, \bibnamefont{and}
  \bibinfo{author}{\bibfnamefont{W.~A.} \bibnamefont{De~Heer}},
  \bibinfo{journal}{Science} \textbf{\bibinfo{volume}{283}},
  \bibinfo{pages}{1513} (\bibinfo{year}{1999}).

\bibitem[{\citenamefont{Wei et~al.}(2001)\citenamefont{Wei, Xie, Dean, and
  Coll}}]{wei-01}
\bibinfo{author}{\bibfnamefont{Y.}~\bibnamefont{Wei}},
  \bibinfo{author}{\bibfnamefont{C.}~\bibnamefont{Xie}},
  \bibinfo{author}{\bibfnamefont{K.~A.} \bibnamefont{Dean}}, \bibnamefont{and}
  \bibinfo{author}{\bibfnamefont{B.~F.} \bibnamefont{Coll}},
  \bibinfo{journal}{Appl. Phys. Lett.} \textbf{\bibinfo{volume}{79}},
  \bibinfo{pages}{4527} (\bibinfo{year}{2001}).

\bibitem[{\citenamefont{Chen et~al.}(2001)\citenamefont{Chen, Saito, Yamada,
  and Matsushige}}]{chen-01}
\bibinfo{author}{\bibfnamefont{X.~Q.} \bibnamefont{Chen}},
  \bibinfo{author}{\bibfnamefont{T.}~\bibnamefont{Saito}},
  \bibinfo{author}{\bibfnamefont{H.}~\bibnamefont{Yamada}}, \bibnamefont{and}
  \bibinfo{author}{\bibfnamefont{K.}~\bibnamefont{Matsushige}},
  \bibinfo{journal}{Appl. Phys. Lett.} \textbf{\bibinfo{volume}{78}},
  \bibinfo{pages}{3714} (\bibinfo{year}{2001}).

\bibitem[{\citenamefont{Kumar et~al.}(2004)\citenamefont{Kumar, Kim, Lee, Song,
  Yang, Nahm, and Suh}}]{kumar-04}
\bibinfo{author}{\bibfnamefont{M.~S.} \bibnamefont{Kumar}},
  \bibinfo{author}{\bibfnamefont{T.~H.} \bibnamefont{Kim}},
  \bibinfo{author}{\bibfnamefont{S.~H.} \bibnamefont{Lee}},
  \bibinfo{author}{\bibfnamefont{S.~M.} \bibnamefont{Song}},
  \bibinfo{author}{\bibfnamefont{J.~W.} \bibnamefont{Yang}},
  \bibinfo{author}{\bibfnamefont{K.~S.} \bibnamefont{Nahm}}, \bibnamefont{and}
  \bibinfo{author}{\bibfnamefont{E.~K.} \bibnamefont{Suh}},
  \bibinfo{journal}{Chem. Phys. Lett.} \textbf{\bibinfo{volume}{383}},
  \bibinfo{pages}{235} (\bibinfo{year}{2004}).

\bibitem[{\citenamefont{Bao et~al.}(2006)\citenamefont{Bao, Zhang, and
  Pan}}]{bao-01}
\bibinfo{author}{\bibfnamefont{Q.}~\bibnamefont{Bao}},
  \bibinfo{author}{\bibfnamefont{H.}~\bibnamefont{Zhang}}, \bibnamefont{and}
  \bibinfo{author}{\bibfnamefont{C.}~\bibnamefont{Pan}},
  \bibinfo{journal}{Appl. Phys. Lett.} \textbf{\bibinfo{volume}{89}},
  \bibinfo{pages}{063124} (\bibinfo{year}{2006}).

\bibitem[{\citenamefont{Langlet et~al.}(2006)\citenamefont{Langlet, Devel, and
  Lambin}}]{langlet-06}
\bibinfo{author}{\bibfnamefont{R.}~\bibnamefont{Langlet}},
  \bibinfo{author}{\bibfnamefont{M.}~\bibnamefont{Devel}}, \bibnamefont{and}
  \bibinfo{author}{\bibfnamefont{P.}~\bibnamefont{Lambin}},
  \bibinfo{journal}{Carbon} \textbf{\bibinfo{volume}{44}},
  \bibinfo{pages}{2883} (\bibinfo{year}{2006}).

\bibitem[{\citenamefont{Wang et~al.}(2007)\citenamefont{Wang, Devel, Langlet,
  and Dulmet}}]{zhaowang-06-01}
\bibinfo{author}{\bibfnamefont{Z.}~\bibnamefont{Wang}},
  \bibinfo{author}{\bibfnamefont{M.}~\bibnamefont{Devel}},
  \bibinfo{author}{\bibfnamefont{R.}~\bibnamefont{Langlet}}, \bibnamefont{and}
  \bibinfo{author}{\bibfnamefont{B.}~\bibnamefont{Dulmet}},
  \bibinfo{journal}{Phys.\ Rev.~B} \textbf{\bibinfo{volume}{75}},
  \bibinfo{pages}{205414} (\bibinfo{year}{2007}).

\bibitem[{\citenamefont{Mayer}(2005)}]{mayer-05-01}
\bibinfo{author}{\bibfnamefont{A.}~\bibnamefont{Mayer}},
  \bibinfo{journal}{Phys.\ Rev.~B} \textbf{\bibinfo{volume}{71}},
  \bibinfo{pages}{235333} (\bibinfo{year}{2005}).

\bibitem[{\citenamefont{Mayer et~al.}(2006)\citenamefont{Mayer, Lambin, and
  Langlet}}]{mayer-06-01}
\bibinfo{author}{\bibfnamefont{A.}~\bibnamefont{Mayer}},
  \bibinfo{author}{\bibfnamefont{P.}~\bibnamefont{Lambin}}, \bibnamefont{and}
  \bibinfo{author}{\bibfnamefont{R.}~\bibnamefont{Langlet}},
  \bibinfo{journal}{Appl. Phys. Lett.} \textbf{\bibinfo{volume}{89}},
  \bibinfo{pages}{063117} (\bibinfo{year}{2006}).

\bibitem[{\citenamefont{Mayer}(2007)}]{mayer-07-01}
\bibinfo{author}{\bibfnamefont{A.}~\bibnamefont{Mayer}},
  \bibinfo{journal}{Phys.\ Rev.~B} \textbf{\bibinfo{volume}{75}},
  \bibinfo{pages}{045407} (\bibinfo{year}{2007}).

\bibitem[{\citenamefont{Stuart et~al.}(2000)\citenamefont{Stuart, Tutein, and
  Harrison}}]{stuart-00}
\bibinfo{author}{\bibfnamefont{S.~J.} \bibnamefont{Stuart}},
  \bibinfo{author}{\bibfnamefont{A.~B.} \bibnamefont{Tutein}},
  \bibnamefont{and} \bibinfo{author}{\bibfnamefont{J.~A.}
  \bibnamefont{Harrison}}, \bibinfo{journal}{J. Chem. Phys.}
  \textbf{\bibinfo{volume}{112}}, \bibinfo{pages}{6472} (\bibinfo{year}{2000}).

\bibitem[{\citenamefont{Brenner}(1990)}]{Brenner-90}
\bibinfo{author}{\bibfnamefont{D.~W.} \bibnamefont{Brenner}},
  \bibinfo{journal}{Phys.\ Rev.~B} \textbf{\bibinfo{volume}{42}},
  \bibinfo{pages}{9458} (\bibinfo{year}{1990}).

\bibitem[{\citenamefont{Guo and Guo}(2003{\natexlab{a}})}]{guo2003}
\bibinfo{author}{\bibfnamefont{Y.}~\bibnamefont{Guo}} \bibnamefont{and}
  \bibinfo{author}{\bibfnamefont{W.}~\bibnamefont{Guo}}, \bibinfo{journal}{J.
  Phys. D: Appl. Phys.} \textbf{\bibinfo{volume}{36}}, \bibinfo{pages}{805}
  (\bibinfo{year}{2003}{\natexlab{a}}).

\bibitem[{\citenamefont{Brenner et~al.}(2002)\citenamefont{Brenner, Shenderova,
  Harrison, Stuart, Ni, and Sinnott}}]{brenner-02}
\bibinfo{author}{\bibfnamefont{D.~W.} \bibnamefont{Brenner}},
  \bibinfo{author}{\bibfnamefont{O.~A.} \bibnamefont{Shenderova}},
  \bibinfo{author}{\bibfnamefont{J.~A.} \bibnamefont{Harrison}},
  \bibinfo{author}{\bibfnamefont{S.~J.} \bibnamefont{Stuart}},
  \bibinfo{author}{\bibfnamefont{B.}~\bibnamefont{Ni}}, \bibnamefont{and}
  \bibinfo{author}{\bibfnamefont{S.~B.} \bibnamefont{Sinnott}},
  \bibinfo{journal}{J. Phys.: Condens. Matter} \textbf{\bibinfo{volume}{14}},
  \bibinfo{pages}{783} (\bibinfo{year}{2002}).

\bibitem[{\citenamefont{Press et~al.}(1992)\citenamefont{Press, Teukolsky,
  Vetterling, and Flannery}}]{nrf77}
\bibinfo{author}{\bibfnamefont{W.~H.} \bibnamefont{Press}},
  \bibinfo{author}{\bibfnamefont{S.~A.} \bibnamefont{Teukolsky}},
  \bibinfo{author}{\bibfnamefont{W.~T.} \bibnamefont{Vetterling}},
  \bibnamefont{and} \bibinfo{author}{\bibfnamefont{B.~P.}
  \bibnamefont{Flannery}}, \emph{\bibinfo{title}{Numerical Recipes in Fortran
  77}} (\bibinfo{publisher}{Cambridge University Press}, \bibinfo{year}{1992}),
  chap.~\bibinfo{chapter}{10}, p. \bibinfo{pages}{413}.

\bibitem[{\citenamefont{Li and Lin}(2006)}]{Li2006}
\bibinfo{author}{\bibfnamefont{T.S.}~\bibnamefont{Li}} \bibnamefont{and}
  \bibinfo{author}{\bibfnamefont{M.F.}~\bibnamefont{Lin}},
  \bibinfo{journal}{Phys.\ Rev.~B} \textbf{\bibinfo{volume}{73}},
  \bibinfo{pages}{075432} (\bibinfo{year}{2006}).

\bibitem[{\citenamefont{Jo et~al.}(2003)\citenamefont{Jo, Tu, Huang, Carnahan,
  Wang, and Ren}}]{Jo2003}
\bibinfo{author}{\bibfnamefont{S.}~\bibnamefont{Jo}},
  \bibinfo{author}{\bibfnamefont{Y.}~\bibnamefont{Tu}},
  \bibinfo{author}{\bibfnamefont{Z.}~\bibnamefont{Huang}},
  \bibinfo{author}{\bibfnamefont{D.}~\bibnamefont{Carnahan}},
  \bibinfo{author}{\bibfnamefont{D.}~\bibnamefont{Wang}}, \bibnamefont{and}
  \bibinfo{author}{\bibfnamefont{Z.}~\bibnamefont{Ren}},
  \bibinfo{journal}{Appl. Phys. Lett.} \textbf{\bibinfo{volume}{82}},
  \bibinfo{pages}{3520} (\bibinfo{year}{2003}).

\bibitem[{\citenamefont{Guo and Guo}(2003{\natexlab{b}})}]{guo-03-02}
\bibinfo{author}{\bibfnamefont{W.}~\bibnamefont{Guo}} \bibnamefont{and}
  \bibinfo{author}{\bibfnamefont{Y.}~\bibnamefont{Guo}},
  \bibinfo{journal}{Phys.\ Rev. Lett.} \textbf{\bibinfo{volume}{91}},
  \bibinfo{pages}{115501} (\bibinfo{year}{2003}{\natexlab{b}}).

\bibitem[{\citenamefont{Treacy et~al.}(1996)\citenamefont{Treacy, Ebbesen, and
  Gibson}}]{treacy-96}
\bibinfo{author}{\bibfnamefont{M.~M.~J.} \bibnamefont{Treacy}},
  \bibinfo{author}{\bibfnamefont{T.~W.} \bibnamefont{Ebbesen}},
  \bibnamefont{and} \bibinfo{author}{\bibfnamefont{J.~M.}
  \bibnamefont{Gibson}}, \bibinfo{journal}{Nature 381, 678}
  \textbf{\bibinfo{volume}{381}}, \bibinfo{pages}{678} (\bibinfo{year}{1996}).

\bibitem[{\citenamefont{Zhang et~al.}(2001)\citenamefont{Zhang, Chang, Cao,
  Wang, Kim, Li, Morris, Yenilmez, Kong, and Dai}}]{zhang-01}
\bibinfo{author}{\bibfnamefont{Y.}~\bibnamefont{Zhang}},
  \bibinfo{author}{\bibfnamefont{A.}~\bibnamefont{Chang}},
  \bibinfo{author}{\bibfnamefont{J.}~\bibnamefont{Cao}},
  \bibinfo{author}{\bibfnamefont{Q.}~\bibnamefont{Wang}},
  \bibinfo{author}{\bibfnamefont{W.}~\bibnamefont{Kim}},
  \bibinfo{author}{\bibfnamefont{Y.}~\bibnamefont{Li}},
  \bibinfo{author}{\bibfnamefont{N.}~\bibnamefont{Morris}},
  \bibinfo{author}{\bibfnamefont{E.}~\bibnamefont{Yenilmez}},
  \bibinfo{author}{\bibfnamefont{J.}~\bibnamefont{Kong}}, \bibnamefont{and}
  \bibinfo{author}{\bibfnamefont{H.}~\bibnamefont{Dai}},
  \bibinfo{journal}{Appl. Phys. Lett.} \textbf{\bibinfo{volume}{79}},
  \bibinfo{pages}{3155} (\bibinfo{year}{2001}).

\bibitem[{\citenamefont{Rochefort et~al.}(1999)\citenamefont{Rochefort,
  Avouris, Lesage, and Salahub}}]{Rochefort1999}
\bibinfo{author}{\bibfnamefont{A.}~\bibnamefont{Rochefort}},
  \bibinfo{author}{\bibfnamefont{P.}~\bibnamefont{Avouris}},
  \bibinfo{author}{\bibfnamefont{F.}~\bibnamefont{Lesage}}, \bibnamefont{and}
  \bibinfo{author}{\bibfnamefont{D.R.}~\bibnamefont{Salahub}},
  \bibinfo{journal}{Phys.\ Rev.~B} \textbf{\bibinfo{volume}{60}},
  \bibinfo{pages}{13824} (\bibinfo{year}{1999}).

\bibitem[{\citenamefont{Nardelli}(1999)}]{Nardelli1999}
\bibinfo{author}{\bibfnamefont{M.B.}~\bibnamefont{Nardelli}},
  \bibinfo{journal}{Phys.\ Rev.~B} \textbf{\bibinfo{volume}{60}},
  \bibinfo{pages}{7828} (\bibinfo{year}{1999}).

\bibitem[{\citenamefont{Coleman et~al.}(2006)\citenamefont{Coleman, Khan, Blau,
  and Gun'ko}}]{coleman-carbon-2006}
\bibinfo{author}{\bibfnamefont{J.~N.} \bibnamefont{Coleman}},
  \bibinfo{author}{\bibfnamefont{U.}~\bibnamefont{Khan}},
  \bibinfo{author}{\bibfnamefont{W.~J.} \bibnamefont{Blau}}, \bibnamefont{and}
  \bibinfo{author}{\bibfnamefont{Y.~K.} \bibnamefont{Gun'ko}},
  \bibinfo{journal}{Carbon} \textbf{\bibinfo{volume}{44}},
  \bibinfo{pages}{1624} (\bibinfo{year}{2006}).

\bibitem[{\citenamefont{Wang et~al.}()\citenamefont{Wang, Devel, Dulmet, and
  Stuart}}]{zhaowang-06}
\bibinfo{author}{\bibfnamefont{Z.}~\bibnamefont{Wang}},
  \bibinfo{author}{\bibfnamefont{M.}~\bibnamefont{Devel}},
  \bibinfo{author}{\bibfnamefont{B.}~\bibnamefont{Dulmet}}, \bibnamefont{and}
  \bibinfo{author}{\bibfnamefont{S.}~\bibnamefont{Stuart}},
  \bibinfo{note}{accepted}.

\bibitem[{\citenamefont{Kozinsky and Marzari}(2006)}]{kozinsky-06}
\bibinfo{author}{\bibfnamefont{B.}~\bibnamefont{Kozinsky}} \bibnamefont{and}
  \bibinfo{author}{\bibfnamefont{N.}~\bibnamefont{Marzari}},
  \bibinfo{journal}{Phys.\ Rev. Lett.} \textbf{\bibinfo{volume}{96}},
  \bibinfo{pages}{166801} (\bibinfo{year}{2006}).

\end{thebibliography}

\end{document}